\documentclass[12pt,preprint]{aastex}  
\newcommand{\mincir}{\raise
-2.truept\hbox{\rlap{\hbox{$\sim$}}\raise5.truept\hbox{$<$}\ }}
\newcommand{\magcir}{\raise
-2.truept\hbox{\rlap{\hbox{$\sim$}}\raise5.truept\hbox{$>$}\ }}
\newcommand{\minmag}{\raise
-2.truept\hbox{\rlap{\hbox{$<$}}\raise6.truept\hbox{$<$}\ }}
\newcommand{\be}{\begin{equation}}
\newcommand{\ee}{\end{equation}}
\newcommand{\ba}{\begin{eqnarray}}
\newcommand{\ea}{\end{esqnarray}}
\newcommand{\brr}{\begin{array}}
\newcommand{\err}{\end{array}}
\newcommand{\bc}{\begin{center}}
\newcommand{\ec}{\end{center}}

\shorttitle{Alignments \& Cluster Dynamics}
\shortauthors{Plionis et al.}

\begin{document}

\title{Galaxy Alignments as a Probe of the Dynamical State of Clusters}

\author{M. Plionis$^{1,2}$, C. Benoist$^{3}$, S. Maurogordato$^{3}$, 
C. Ferrari$^{3}$, S. Basilakos$^{2}$}
\affil{$^{1}$ Instituto Nacional de Astrofisica, Optica y Electronica (INAOE)
 Apartado Postal 51 y 216, 72000, Puebla, Pue., Mexico 
\\
$^{2}$Institute of Astronomy \& Astrophysics, National Observatory of
Athens, I.Metaxa \& B.Pavlou, P.Penteli 152 36, Athens, Greece
\\
$^{3}$ Observatoire de la Cote d'Azur, CERGA, BP229, Nice, cedex 4, France}

\begin{abstract}

We present indications, based on a sample of 303 Abell clusters, 
for a relation between the dynamical state of 
clusters and the alignments of galaxy members with their parent cluster 
major axis orientation as well as with 
the large scale environment within which the clusters are embedded.
The statistical results are complemented with a deep, wide-field case
study of galaxy alignments in the cluster A521, which is characterised
by multiple merging events (Maurogrdato et al 2000, Ferrari et
al. 2003) and whose galaxy members show a strong alignment signal 
out to $\sim 5 \; h^{-1}$ Mpc.
Our results show that galaxy alignments appear 
to be stronger the more dynamically 
young is the cluster, especially when found in high-density environments.
This relation complements the recently found {``Cluster Substructure - 
Alignment Connection''} (Plionis \& Basilakos 2002) 
by which dynamically young clusters, found in high-density
environments, show stronger cluster-cluster alignments.
\end{abstract}


\section{Introduction}

A well established alignment effect is that between the orientation of
nearby clusters and between the orientation of the BCG (or cD) and
that of their parent cluster (cf. Sastry 1968; Binggeli 1982; Carter
\& Metcalfe 1980; Struble 1990; West 1989; West 1994; Plionis 1994;
Fuller, West \& Bridges 1999; Kim et al. 2001; Chambers, Melott,
Miller 2002).  Furthermore, recent evidence has shown that
substructures in clusters are aligned with the cluster orientation as
well as along the large-scale filamentary structures within which they
are embedded (cf. West, Jones \& Forman 1995, Plionis \& Basilakos
2002, Plionis 2001).  Analytical (cf. Bond 1987) and numerical work 
(cf. West, Villumsen, Dekel 1991; van Haarlem \& Van de Weygaert 1993, 
Tormen 1997, Onuora \& Thomas 1999; Splinter et al. 1998; 
Falthebacher et al. 2002) show that
cluster-cluster and substructure-cluster alignments occur naturally in
hierarchical clustering models of structure formation, like the
CDM. This fact has been explained as the result of an interesting
property of Gaussian random fields that occurs for a wide range of
initial conditions and which is the "cross-talk" between density
fluctuations on different scales. This property is apparently also the
cause of the observed filamentariness observed not only in "pancake"
models but also in hierarchical models of structure formation; the
strength of the effect, however, differs from model to model.

Recent observational evidences point also to the so called 
``Cluster Alignment - 
Substructure Connection'' by which the dynamically young clusters show a 
stronger tendency to be aligned (Plionis \& Basilakos 2002) 
while they are also located preferentially in high-density environments 
(Plionis \& Basilakos 2002; Sch\"{u}ecker et al 2001). Such alignment 
effects could be imprinted also in smaller scales; the scales of 
individual galaxies.
For example, if galaxies formed after the collapse of their parent 
cluster, then the anisotropic initial conditions could be imprinted 
in member galaxy orientations. 
In the hierarchical structure formation models, galaxy
alignments could originate from a combination of different mechanisms,
for example as a result of the parent cluster tidal
field (cf. Salvador-Sole \& Solanes 1993; Usami \& Fujimoto 1997; but
see also Barnes \& Efstathiou 1987), a possibility supported 
also by the correlation found between disk galaxy spin axes and the 
local tidal shear field (Lee \& Pen 2002), 
and/or if the galaxy-galaxy interactions occur in a
preferred direction, for example along the primordial large-scale
filamentary structure within which the protocluster is embedded. This
anisotropic merger scenario of West (1994) has provided an interesting
explanation of the observed strong alignment of the brightest cluster
galaxy (BCG) not only with the cluster position angle but also with the
near large-scale structures. 

In clusters, any primordial galaxy alignments effect should be
severely damped by violent relaxation, by the exchange of angular
momentum in multiple galaxy encounters that occur in the dense
cluster environment over a Hubble time (cf. Coutts 1996) and even by
secondary infall (Quinn \& Binney 1992).  Therefore, other than the
alignment of the BCG with its parent cluster that could be expected
(cf. Struble 1990; West 1994; Kim et al. 2001), 
it seems secure to say that in highly relaxed clusters,
where there has been sufficient time to mix the phases, one should not
expect to observe any significant primordial galaxy alignment, 
even if they did originally exist. Therefore,
the existence or not of galaxy alignments in and around clusters could
be an indication of their dynamical state.
% in which case one can hope
%to observe galaxy alignments either in low-density environments, like
%in the outskirts of clusters, or in unrelaxed, dynamically young
%clusters.
The possible existence of such alignments, intrinsic in nature, 
is important also for their effect on measures
of weak lensing, where ellipticity correlations of
background galaxies are expected to arise from the lensing 
of foreground large mass inhomogeneities (cf. Bartelmann \& Schneider
2001). Considerable
effort has been devoted, recently, in attempts to quantify and
disentangle these effects (cf. Catelan, Kamionkowski \& Blandford 1996; 
King \& Schneider 2003; Heymans \& Heavens 2003).

Observational efforts to detect intrinsic galaxy alignments in clusters or
superclusters has not reached a clear consensus. 
For example, Dekel (1985) using the UGC and ESO-Uppsala galaxy
catalogues concluded that no significant galaxy alignments exist in
clusters.  
However, earlier studies of Adams, Storm \& Storm (1980)
had found alignments, although not very significant, in A999 and
A2197. Djorgovski (1983) found galaxy alignments in the Coma cluster,
an effect more prominent for red galaxies, while van Kampen \& Rhee
(1990) and Trevese, Cirimele \& Flin (1992) analysing the 
10-20 brightest galaxies in a large number of Abell clusters 
(Abell 1958; Abell, Corwin \& Olowin 1989) found no galaxy
alignments other than between the BCG and its parent cluster
orientation (see Djorgovski 1987; Cabanela \& Aldering 1998 and 
references therein for reviews of early results). 

One should be very careful in how to interpret such conflicting
results. It should be noted that systematic biases, like PSF inhomogeneities
(cf. Brown et al 2002), and projection effects can to mask any 
true alignment signal. 
For example, the projection of foreground galaxies along the line of 
sight of a cluster as well as the projection on the plane of the sky 
 of member galaxies, always work in the direction of smearing alignments. 
On the other side, telescope tracking problems can create false alignments.
Such problems, if not modeled adequately, could potentially 
introduce artificial galaxy-galaxy alignments, and thus hamper attempts to
measure weak shear induced by gravitational lensing, but cannot
create a false alignment between galaxies and their parent cluster
orientation, since the later is estimated from the distribution of
the galaxy member positions.

Therefore, the question whether galaxy alignments exist in clusters
and if they do then under which circumstances do they appear, is still
unanswered.
% and careful analysis, taking into account all possible
%systematic effects, is due. 
In this paper we present a statistical
analysis of galaxy alignments in 303 Abell clusters (Abell 1958; Abell, 
Corwin \& Olowin 1989), based on their
POSS scans and a {\em SExtractor} (Bertin \& Arnouts 1996)
identification of the galaxy
distribution within a radius of 1 $h^{-1}$ Mpc from the cluster
centre. These statistical results are complemented with a deep, wide-field
case study of galaxy alignments within $\sim 5 \; h^{-1}$ Mpc of the 
cluster A521, which is a relatively distant ($z \sim 0.25$), 
richness class 1 Abell cluster 
with clear evidence of multiple merging events (Maurogordato et al. 2000; 
Arnaud et al. 2000; Ferrari et al. 2003).

\section{Galaxy Alignments in Abell Clusters}

We have analysed the POSS scans in the
direction of 303 Abell clusters, all of which have an independent
determination of their position angle (from Plionis 1994), while 168
also have velocity dispersion measurements (Struble \& Rood 1999;
De Propris et al. 2001). Our sample, which was defined on the basis of
availability of position angle measurements in Plionis (1994), has
$\langle z \rangle \simeq 0.07$ and $m_{10}<17$.

An area of radius spanning from $\sim 17^{'}$ to $40^{'}$ was used
so as to correspond to $\sim 1 \; h^{-1}$ Mpc at the distance of the
clusters. In the cases of nearby clusters ($z \mincir 0.045$) the
corresponding radius is $< 1 \; h^{-1}$ Mpc.  An interactive procedure
was followed in which the parameters of the {\em SExtractor}
identification were varied so as to maximise the number of detected
objects, while not breaking-up the brightest galaxies into
subcomponents. Although the {\em SExtractor} star-galaxy separation is
quite efficient, it fails to correctly separate stars from galaxies
for saturated objects and at the faint luminosity end. Bright stars 
were therefore excluded by hand. Similarly, satellite tracks and
other noise signal were excluded after careful visual inspection
of each field. 

The total number of Sextracted galaxies is $\sim 5\times 10^{5}$ out of
which the majority covers a very small surface area and thus have
unreliable classification.
In the presence of no systematic effects, related 
to the size of the galaxies, we expect that the mean ellipticity should be
independent of the galaxy image size. In figure 1 we plot the derived 
galaxy ellipticity as a function of the number of pixels per galaxy. 
It is evident that for 
$N_{\rm pix}\lesssim 30-35$ there is a systematic offset of the mean
ellipticity to higher values. Therefore, in order to avoid
biases, having to do with determining galaxy shape parameters using
small number of pixels and the
inefficiency of the star-galaxy separation at lower surface
brightness, we choose to analyse only the higher 5\% quantile
of the galaxy size distribution (estimated independently in each
field). The selection of this limit was based on a further trial and error
approach in which the distribution of galaxy position angles was
tested for homogeneity (see tests below) and it is such that always 
$N_{\rm pix}>30-35$.

Furthermore, we have tested, for the above size limits, the
ellipticities of the unsaturated stellar images in our fields, for which 
deviations from zero should reflect possible PSF anisotropies.
We find that the stellar ellipticity
frequency distribution peaks at $\epsilon \simeq 0.05$ and has an exponential
drop thereafter (figure 2). We have verified by visual inspection of a
random subset of our fields that the major contribution
to the high-$\epsilon$ tail of the stellar ellipticity distribution 
comes from misclassified galaxies, double stars as well as background
projections near the stellar images. 

Therefore, from the above analysis and 
in order to avoid ill definitions of position angles (due
to near sphericity) and possible PSF anisotropies while not throwing away
too many galaxies, since a considerable fraction of ellipticals have 
relatively small ellipticities, we feel secure to use, in our study,
galaxies that have ellipticities $>0.1$. In any case, we expect that
any residual anisotropy that still remains in our sample will be
identified in the Fourier-analysis test, presented below, but
should bare no correlation with the parent cluster orientations
and thus should not introduce any artificial alignments.
These selection procedures reduce the total number of analysed
galaxies to 15560, with no galaxies having an area
smaller than $\sim 10^{''} \times 10^{''}$ (ie., $N_{\rm pix}>35$). 
%In figure 2 we show the disribution of the number
%of galaxies per cluster field finally analysed.

The statistical analysis of alignments between galaxy and cluster
orientations depends on defining the relative orientation of the
galaxies with respect to the cluster, 
\be
\phi_{i,c}\equiv |\theta_i -\theta_c| \;\;.
\ee  
where $\theta_i$ and $\theta_c$ are the position angles of the
galaxies and cluster, respectively.
In an isotropic distribution we will have $\langle
\phi_{i,c} \rangle \simeq 45^{\circ}$.  A significant deviation from
this value would be an indication of an anisotropic distribution which can
be quantified by (Struble \& Peebles 1985):
\begin{equation}\label{eq:alin}
\delta=\sum_i \frac{\phi_{i,c}}{N}-45 \;\;.
\end{equation}
In an isotropic distribution we will have $\langle \delta \rangle
\simeq 0$, while the standard deviation is given by
$\sigma=90/\sqrt{12 N}$.  A negative value of $\delta$ would indicate
alignment and a positive one misalignment. 

%Furthermore, we exclude from our
%analysis the cD or BCG of each cluster, to avoid contaminating our
%signal by their already well-established alignment effect (Struble ??,
%West ???).

\subsection{Tests for systematics directional biases}
Before we analyse the possible galaxy-cluster alignments, we test
whether the galaxy position angles, $\theta_i$, in our final galaxy 
sample, have any significant directional bias. 
We quantify possible deviations from isotropy by a variety of methods:
\begin{itemize}
\item estimating the Fourier transform of the galaxy position angles
(cf. Struble \& Peebles 1985):
\begin{equation}\label{eq:f1}
C_{n} = (2/N)^{1/2} \sum \cos 2n\theta_i\;\;\;\;\;\;\;\;
S_{n} = (2/N)^{1/2} \sum \sin 2n\theta_i \;.
\end{equation}
If the $\theta_i$'s are uniformly distributed between $0^{\circ}$ and
$180^{\circ}$, then for large N 
both $C_{n}$ and $S_{n}$ have zero mean and unit
standard deviation. This test measures the tendency of the position
angles to cluster around some particular values. For example,
the $C_1$ and $S_1$ components measure deviations in the
$90^{\circ}$, $180^{\circ}$ and $45^{\circ}$, $135^{\circ}$  
directions respectively. 
%Therefore, these components are ideal to
%check for possible star contamination of the galaxy sample since the
%diffraction pattern will produce fake elongations along such directions.
Indeed, if we find Fourier components with values $\gtrsim 3$ 
this would imply systematic directional bias at a $>3 \sigma$ level. 
\item applying three statistical tests on the distribution of $\theta$'s
to test whether they could originate from an isotropic parent distribution 
of position angles. These are the classical $\chi^2$ (parametric), the
Kolmogorov-Smirnov (non-parametric) and the Rayleigh tests (cf. Mardia
1972). The later estimates the parameter: 
\be
 R_2 = \frac{1}{\sqrt{2 N}} \left( S_1^2 + C_1^2 \right)^{1/2}\;\;\;.
\ee
For large N we have
that $2 N R_2^2 \approx \chi^2$ for 2 degrees of freedom and the
probability $P_{R_2}$ that a value of $R_2$ exceeds a critical value
is given in Mardia (1972).  
\end{itemize}

In figure 3 we present
the distribution of galaxy position angles for our total sample as
well as for the 10 largest galaxies (excluding the BCG)
in each cluster field.
The distribution appears flat with no directional bias, 
 as quantified also by the different statistical tests, presented
in table 1, where we also list
the values of the first Fourier moments for 
the different galaxy samples. We do find $|C_{1,2,3,4}|,
|S_{1,2,3,4}|\lesssim 2.3$, indicating the existence of no significant
directional bias. However, had we imposed no galaxy size limit
we would have
found significant deviations from isotropy with very large values of
the first moments due to inefficient star-galaxy separation 
(we deduce this from the fact that we find the same
anisotropy pattern when we analyse the stars in our fields).

The above analysis is based in treating all galaxies together,
irrespectively of the cluster field to which they belong. This
does not exclude the possibility that
some cluster fields maybe severely affected by biases, a possibility
that could be hidden when treating all the galaxies together. To this
end we calculate the Fourier moments independently in each field. If
there is no bias then the distribution, over the 303 fields, 
of the individual Fourier
moments should be a Gaussian with zero mean and unit standard
deviation. In figure 4 we plot their distributions and
the corresponding zero-mean Gaussians. We find $\langle C_1 \rangle =-0.09$,
$\sigma_{c_1}=1.20$ and $\langle S_1 \rangle =0.00$,
$\sigma_{s_1}=1.03$. A $\chi^{2}$ test shows that the
distributions are statistically equivalent to zero-mean Gaussians at
a $>0.1$ significance level
(we find similar results for the higher moments as well). Therefore it
seems that the small excess of $C_1$'s at negative values is not very
significant although it is responsible for the small negative shift of the
$\langle C_1 \rangle$ and for the 20\% excess of its variance. 
Indeed if we exclude the 7 cluster fields with $|C_1|>3$ we
reduce the overall value of $C_1$ (see table 1) from -2.3 to -0.9 and the
mean over the 296 cluster fields to $\langle C_1 \rangle \simeq
-0.02$. However, our alignment results, presented below, remain unchanged
when excluding these fields.

\subsection{Results}
In order to maximise the alignment signal to noise ratio we have
stacked all cluster images after rotating them so as to have a common
orientation along their determined position angles. We then estimate
the misalignment angle between every galaxy and the common cluster
orientation. In order to identify where does the possible alignment
signal come from, we apply our test to different subsamples of
galaxies. One is only the Brightest Cluster galaxy (BCG), another is
the next 10 largest galaxies (10LGs) 
and finally all the galaxies but the BCG and 10LGs. 
The alignment results for different subsamples are presented in table 2.
We find that the over all 15560 galaxies in the 303
clusters, alignment signal is quite weak but significant ($\delta=-0.9$,
$\sigma=0.2$). Confining ourselves to the 10LGs
we find a stronger alignment signal. 
Finally and as expected, we find a very strong and
statistically significant
alignment signal between the BCG's and the clusters position angles.  

\subsubsection{Statistical quantification}
In figure 5 we present the distribution of misalignment
angles, $\phi$ for the BCG and 10LGs case.
We performed a $\chi^2$ test with 5 degrees of freedom
to assess whether the observed distribution
could be drawn from an isotropic one. Furthermore we also applied
a non-parametric Kolmogorov-Smirnov (KS) test and compared
the un-binned $\phi$-values with the isotropic case. The
results of these tests are presented in Table 2. There are
significant deviations from isotropy for the sample
containing all the galaxies as well as for the
BCG and the 10LGs samples. Note however, that when excluding the
BCG's and 10LG's no significant deviation from isotropy is found (last
row in table 2). For fainter galaxies the combination of projection 
effects, which tend to smear out any existing alignment, and shape parameter 
uncertainties, could hide an existing alignment effect and thus we cannot
derive any strong conclusion from our present analysis.
Unfortunately, we do not have colour information 
in order to attempt to decontaminate the cluster fields 
by selecting the red-sequence galaxies, as we will do in our 
case study of A521.

Also note that the joint probability
of the observing the  alignments of the
BCG and 10LGS samples should increase if one
considers (a) the two samples as being independent of each other
and (b) that the deviation from isotropy, in both, is toward the same
(alignment) direction.

\subsubsection{Dependence on environment}
We investigate further whether there is any correlation of the
alignment signal with the environment.
%and/or dynamical youth (as indicated
%for example from high galaxy velocity dispersion). 
It is already known that clusters in high-density environments show
stronger cluster-cluster alignments and more substructure than in
low-density ones, as would be expected if such clusters are still
accreting matter along the large-scale filamentary structures in which
they are embedded (cf. Plionis \& Basilakos 2002; Sch\"{u}ecker et
al. 2001). We have therefore found which of the clusters in the
subsample analysed do belong to superclusters.  To this end, we have
performed a percolation analysis to identify superclusters in the
whole Abell sample with a magnitude cut $m_{10}<17$, corresponding to
a volume limited sample out to $z \sim 0.09$.  We do find a weak tendency
of increasing galaxy-cluster alignment signal in clusters belonging to
higher overdensity superclusters. For example using clusters in
superclusters found with a percolation radius of 15 $h^{-1}$ Mpc one
finds ($\delta_{\rm BCG}=-11.3$, $\sigma_{\rm BCG}=2.2$ and
$\delta_{\rm 10LG}=-2.1$, $\sigma_{\rm 10LG}=0.7$). This tendency, weak as 
it may be, suggests a correlation between the galaxy-cluster alignment 
signal and the
cluster dynamical state, since in high density environments, clusters
are expected to be dynamically active as discussed previously.

\subsubsection{Dependence on cluster dynamical state}
We want to test the alignment behaviour of the cluster high
velocity dispersion tail, expected to correspond to highly virialised
massive clusters, since the virialisation process should
have mixed the orientation of galaxies members, and thus weak or no galaxy
alignments are expected. For this purpose we have analysed for alignments
a subsample of 132 clusters that have estimated $\sigma_v$ with more
than 10 galaxies per cluster.

For the 32 clusters in our sample with $\sigma_v\ge 900$ km/sec, we obtain 
$\delta_{\rm 10LG} \simeq -6.1$, $\sigma_{\rm 10LG} \simeq 1.5$ with 
a distribution of
misalignment angles being consistent with the isotropic case at a 
significance level of $P_{>\chi^2} \sim 2 \times 10^{-4}$ 
and $P_{KS} \lesssim 10^{-4}$.
%, a signal which
%increases to $\delta=-5.8$, $\sigma=1.7$, 
%$P_{>\chi^2}< 10^{-5}$ and $P_{KS} \sim 0.04$) for clusters in
%superclusters with $r_{\rm perc}=10 \; h^{-1}$ Mpc.
Correspondingly for the remaining 100 clusters of our
sample with $\sigma_v< 900$ km/sec we obtain a weak alignment
signal $\delta_{\rm 10LG} \simeq -1.5$,
$\sigma_{\rm 10LG} \simeq 0.8$.  These results show a clear correlation of
the alignment effect with velocity dispersion, which is in the opposite
direction expected if the high dispersion clusters were indeed
highly relaxed clusters.

If we further confine our analysis in clusters belonging to dense
superclusters (percolation radius of $r_{\rm perc}=15 \; h^{-1}$ Mpc) and
with $\sigma\ge 900$ km/sec (in total 19 clusters) we obtain an even
stronger alignment signal: $\delta_{\rm 10b}=-7.4$, $\sigma_{\rm 10LG}=
1.9$, higher than in both cases taken separately. In figure 6 we present the 
distribution of misalignment angles for the latter case (red hatched 
histograms) and for the clusters with $\sigma_v< 900$ found in any environment.
The excess alignment signal for the high-$\sigma$ clusters in dense environment
is evident. Their probability of consistency with an isotropic distribution is
$P_{KS}< 0.002$ for both the BCG and 10LGs distributions. 
 
%We have then examined in details this subsample of Abell clusters.
% which Abell numbers are listed in Table XXX. 

At this point, the high density environment of these clusters makes us 
suspect that their high velocity dispersion does not reflect high mass
virialised clusters but rather the result of subclustering effects, 
(for example, due to the large relative velocities of the different
sub-clumps), so that these objects would be in reality dynamically
young clusters. We
have then been looking in the literature for each individual cluster
if signatures of dynamical youth can be detected in their X-ray, optical or
radio images. We have found that for 12 
(A1228, A2061, A399/A401, A1736, A1775, A2065, A401, A426, A754, A2256
and A85) of the 19 clusters, at least one
signature does exist, for a further 3 we have not found any 
relevant information while only 4 appear to be virialised objects.

In detail, A1228, A2061, A399/A401, A1736 and A1775 show a multi-peaked
velocity distribution, A2065, A401, A426 and A754 show features in the
X-Ray temperature maps, A2256 shows sub-clustering in the optical
density map. Moreover, A1190, A2256 and A754 show radio halo and relics,
which are thought to be related to merging events (cf. Feretti \&
Giovanninni 1996, Feretti 2001). 
Finally, A85 presents signatures of merging in optical
(presence of a foreground group in the velocity distribution),
X-ray (double configuration of the isointensity contour map, with a primary
and a secondary cluster) and in radio bands (presence of a radio relic).

Since there is a triple correlation between alignments, velocity
dispersion and superclustering, we present in figure 7 the alignment
signal as a function of cluster velocity dispersion and percolation
radius (for the 132 clusters of our sample that have measured
$\sigma_v$ and $N_z>10$).
 The alignment signal is represented as a grey scale, with
increasing brightness meaning more alignments, as indicated by the
printed values of $\langle \delta \rangle$, while the contour lines
engulf regions of some significance level (the continuous and broken lines
represent the 3 and 2.5 $\sigma$ levels respectively). It is evident
that for the 10LG case there is a significant trend of increasing
alignment signal both with increasing $\sigma_v$ and supercluster 
percolation radius. It is also interesting that for the BCG's there seems
to be a dependence of their alignment, with their parent cluster
orientation, mostly to the environment, almost independent of their
velocity dispersion. This could indicate a different (or ``special'') 
formation process for the BCG, with respect to the other luminous
cluster galaxies (cf. West 1994).

%Note that the relatively lower alignment signal 
%for $r_{\rm perc}\lesssim 10 \; h^{-1}$ Mpc is due to small number
%statistics, as indicated also by the low-significance (it lies outside
%the 2.5$\sigma$ level).

Our results indicate that cluster alignments are stronger in clusters
located in a high density environment and characterised by a
high-velocity dispersion. Since we
already have independent indications that clusters in high-density
environments show significantly more substructure 
(Plionis \& Basilakos 2002; Sch\"{u}ecker et al. 2001),
our results suggest that the strongest alignment effects have been 
detected in dynamically young clusters, ie., characterised by both a
strong subclustering and a high-velocity dispersion.

\section{A case study: the dynamical active Abell 521}

The galaxy cluster A521 is a dynamically very active cluster with
multiple evidences of merging both in optical and X-Ray data. In
particular, its very high velocity dispersion of $\sim 1300$ km/s
(Maurogordato et al. 2000) has been explained as the result of various
merging events (Ferrari et al. 2003). The main component of the cluster
shows a clear North-West/South-East direction, both in X-ray and in
optical, which is the main direction of the undergoing merger event,
but also a high density ridge in the core region perpendicular to
the main direction, suspected to result from an older merger.  The
cluster is embedded in a dense large-scale environment (Arnaud et al. 
2000). At the light of the results of the previous section, 
this high velocity dispersion and highly substructured cluster embedded
in high density environment should be an excellent target for
detecting galaxy alignments.

Part of a larger program of a multi-wavelength follow up of merging
clusters, Abell 521 has been covered with B- and I-band imaging
obtained with the CFHT12k camera over a field of 45$\times$30
arcmin$^2$ centred on the cluster.  The galaxy catalogue (positions,
magnitudes and shape parameters) was built with {\em SExtractor}.
From the analysis of the I$\times$(B-I) colour-magnitude
diagram we have extracted all the galaxies lying on the red sequence
region allowing to select the most likely early type galaxies
belonging to the cluster (see Ferrari et al. 2003). Such a filtering 
of the galaxy catalog is very important since
foreground or background galaxies can significantly smear out 
any true alignment effect. The alignment analysis presented in the 
present paper has been performed
using the shape parameters measured in the B-band image leading to a
more precise measurement of the position angles than in the I-band
where the light profiles of the galaxies are more peaked. In figure 8 we 
show the stellar ellipticity frequency distribution which peaks at
$\epsilon \simeq 0.03$ and drops exponentially thereafter. Therefore,
to avoid the gross effects of possible 
PSF inhomogeneities and to ensure a robust position angle determination
(since it is ill defined for nearly spherical objects),
we have imposed a minimum ellipticity of $\epsilon=0.1$.

\subsection{A521 Shape Parameters}
The shape parameters of the cluster (ellipticity and position angle) were
estimated using the familiar moments of inertia method applied on the smoothed 
or the discrete galaxy distribution (cf. Basilakos,
Plionis \& Maddox 2000 for such an application).  
First, all galaxies
within an initial small radius around the cluster cD or X-ray centre
(Arnaud et al. 2000) are used to define the initial value of the
cluster shape parameters.  Then, the next nearest galaxy is added to
the initial group and the shape is recalculated and thus we obtain the
cluster shape parameters as a function of cluster-centric distance.
%This method has been found to provide, in the unavoidable presence of
%foreground and background galaxies, biased low values of the cluster
%ellipticity but not of the position angle (Basilakos et al. 2000).  
Within 1 $h^{-1}$ Mpc of the central
cD galaxy the cluster position angle and ellipticity have value of:
$$\theta_{c} \approx 140^{\circ} \pm 15^{\circ} \;\;\;\;\;\ \&
\;\;\;\;\;\ \epsilon \approx 0.45 \pm 0.1 \;\;,$$ 
where the uncertainty is derived from the
fluctuations around the mean by varying the limiting magnitude or by
weighting each galaxy either by its $L_{b}$ or by unity. 
%Note that the 
%ellipticity of A521 was determined by the {\em smoothing} approach of 
%Basilakos, Plionis \& Maddox (2000), which does not underestimate its value.

\subsection{Galaxy member orientations}
In figure 9 we present the galaxy position angles 
for different magnitude slices and for two limiting distances from 
the cluster centre, parametrised by $m_{*}$ which 
corresponds to the knee of the B-band optical galaxy luminosity function
%: $m_{*}=M_{*}+5 \log \left( \frac{c z}{H_{\circ}} \right)+25+3 z \;,$ 
where $M_{*}\approx-19.7$ and therefore $m_{*} \approx 20.8$. 
There is an evident
concentration around $120^{\circ} \sim 160^{\circ}$ in all magnitude
bins, although at fainter magnitudes there is a slight shift toward
the lower end of this range. This can be quantified by estimating the
preferred galaxy orientation of the galaxy position angle distribution
using the fundamental harmonics ($n=1$) of their Fourier transform
(cf. Djorgovski 1983):
\begin{equation}
\theta_{\rm pr} = \frac{1}{2} \arctan \left(S_1/C_1 \right) \;\;.
\end{equation}
The values of $\theta_{\rm pr}$ for the different magnitude slices within
1.5 and 2.5 $h^{-1}$ Mpc from the cluster centre, are
reported in table 3 and it is evident that they are in good agreement
with the cluster position angle which also coincides with the
direction ($\phi\simeq 138^{\circ}$) to the nearest Abell cluster 
neighbour (Abell 517). These results are robust to changes of the limiting 
galaxy ellipticity as can be seen in table 3.
Furthermore, it is extremely interesting that the same galaxy
preferred orientation is evident within 1.5 and also within 2.5
$h^{-1}$ Mpc from the cluster central cD galaxy.
% an unprecedented
%coherence in the anisotropic galaxy position angle distribution,
%extending to the outer environment of the cluster.
The fact that the preferred galaxy orientation, $\theta_{\rm pr}$,
coincides with the orientations of the cluster major axis, of the
brightest cluster galaxies and with the direction 
to the nearest cluster neighbour is a strong indication
that the anisotropic orientation distribution is not due to possible
systematics and strongly supports the anisotropic merger structure formation
scenario.

In table 3 we also present the outcome of the statistical tests used to
assess the significance of our results. It is evident that
the brightest and faintest slices present results
significantly different from the
Poisson expectations, while the intermediate slice shows low
significant difference. This however could be because in this
magnitude range, there appears to be an excess of galaxies with
position angles at $\sim 30^{\circ}$, which coincides with the
direction of the ridge structure {\em S1} of Arnaud et al. (2000) and
which is mostly evident in this magnitude range.
Note that if we use all the galaxy position angles down to $m\sim 24$
all three tests show a significant difference from the Poisson
expectation at a 99.999\% level.

\subsection{A521 groups orientation}
Since the matter distribution outside the cluster itself could be 
anisotropically distributed, not necessarily following the 
orientation indicated by the cluster major-axis, we have estimated the 
preferred orientation of the high-density regions
as well as the galaxy major axis preferred
orientation in independent shells of $1 \;h^{-1}$ Mpc width each.
%whether the same 
%anisotropic distribution observed in the galaxy orientations 
%is reflected also in
To this end we estimate the major axis 
orientation of the projected high-density excursion set (which correspond 
to groups of galaxies and different substructures) by Gaussianly smoothing
the projected galaxy
distribution on a $N\times N$ grid (with $N$ ranging from 300 to 400)
and measuring the position angles of dense groups ($\rho/\langle \rho
\rangle > 14$) containing more than 12 cells (surface area $\magcir
2700$ kpc$^2$).  

We find that the group orientation
distribution closely follows that of the individual galaxies with an
almost identical preferred orientation angle, $\theta_{\rm pr}$.  This
is clearly seen in figure 10 were we present the histogram of the group
and galaxies position angles (red and black histogram respectively),
for two cluster-centric distances (one within the nominal radius of 2
$h^{-1}$ Mpc and one between 2 and 5 $h^{-1}$ Mpc; were the last group
is found).  Note that, in order to detect groups in the outer shell, we
have reduced the density threshold to $\rho/\langle \rho \rangle >
11$. 
In table 4 we present the preferred orientations of groups and of galaxies 
in the different shells, their misalignment angle and
the probability that this misalignment angle is due to chance. 
The $>30^{\circ}$ misalignment angle between the galaxy and
group preferred orientations in the 3$^{rd}$ and 4$^{th}$ bin
reduces to $< 10^{\circ}$ if we apply
our analysis to a lower limiting magnitude ($m_{\rm b,lim}\sim 23.8$),
since we do expect a higher fraction of projection contamination from
apparently faint galaxies outside the dense cluster region. 

%The orientation 
%coherence is evident out to $\sim 5 \; h^{-1}$ Mpc from the cluster centre.
\subsection{Outcome}
We therefore conclude that the dynamically very young and in the phase
of merging Abell 521 shows significant red-sequence galaxy major-axis
alignments, within the cluster but also in its nearby environment
($\mincir 5 \; h^{-1}$ Mpc), with 
{\em (a)} the cluster major-axis orientation,
{\em (b)} the groups/substructures preferred orientation and
{\em (c)} with the direction of the nearest Abell cluster (A517).

\section{Summary}
We have set out to investigate whether galaxies in clusters show any
alignment of their major axis with the cluster orientation and whether
such an effect is related to the cluster environment and dynamical state. 
We tackle this question by using a statistical approach, 
analysing the POSS scans of 303 Abell clusters ($\langle z \rangle \simeq
0.07$, $m_{10}<17$) and a case study of a highly dynamically active 
cluster (A521) using multicolour, wide-field, deep imaging data.

In order not to contaminate our statistical results with the
well-established BCG-cluster alignment we have excluded these galaxies
from our analysis. We find that clusters indeed show significant
bright galaxy major axis alignments with the cluster position angle but only
in high-density environments (superclusters). Furthermore, the
alignment signal is stronger in higher velocity dispersion clusters.
Moreover, we have also found in our sample that the high velocity 
dispersion clusters, in high density environments, 
appear to be dynamically young, merging clusters, 
a fact which points strongly in the direction of galaxy alignments being 
correlated with their parent cluster dynamical youth.

Similarly, in our detailed analysis of the dynamically very active
Abell 521 we have found significant red-sequence galaxy alignments out
to $\sim 5 \;h^{-1}$ Mpc from cluster core. The alignments are evident
within the cluster as well in the outer environment and the preferred
galaxy orientation is closely related to the mass distribution orientation, 
as well as with the direction of the nearest Abell cluster (A517).

These results provide evidence that significant galaxy alignments are
present in dynamically young clusters belonging in superclusters, 
that also extend to the large-scale environment around clusters. 
%being probably associated with the anisotropic matter distribution 
%from which the clusters accrete matter. 
Furthermore, these results lead also to
the necessity of a possible revision of all those cluster 
weak lensing studies done in only one pass band, in which case
it is impossible to separate background objects from the cluster members.

\acknowledgments
{\small
We would like to thank the secondary school students of the 4$th$ and 
5$th$ Astrophysics Summer School for their dedication and help in analysing 
part of the Abell cluster images used in this study. This work has been 
supported by the 
Greek-French scientific bilateral program (2002-2003) entitled
{\em Study of the formation of large-scale structure using optical and 
X-ray data'}.}

\begin{figure}
\plotone{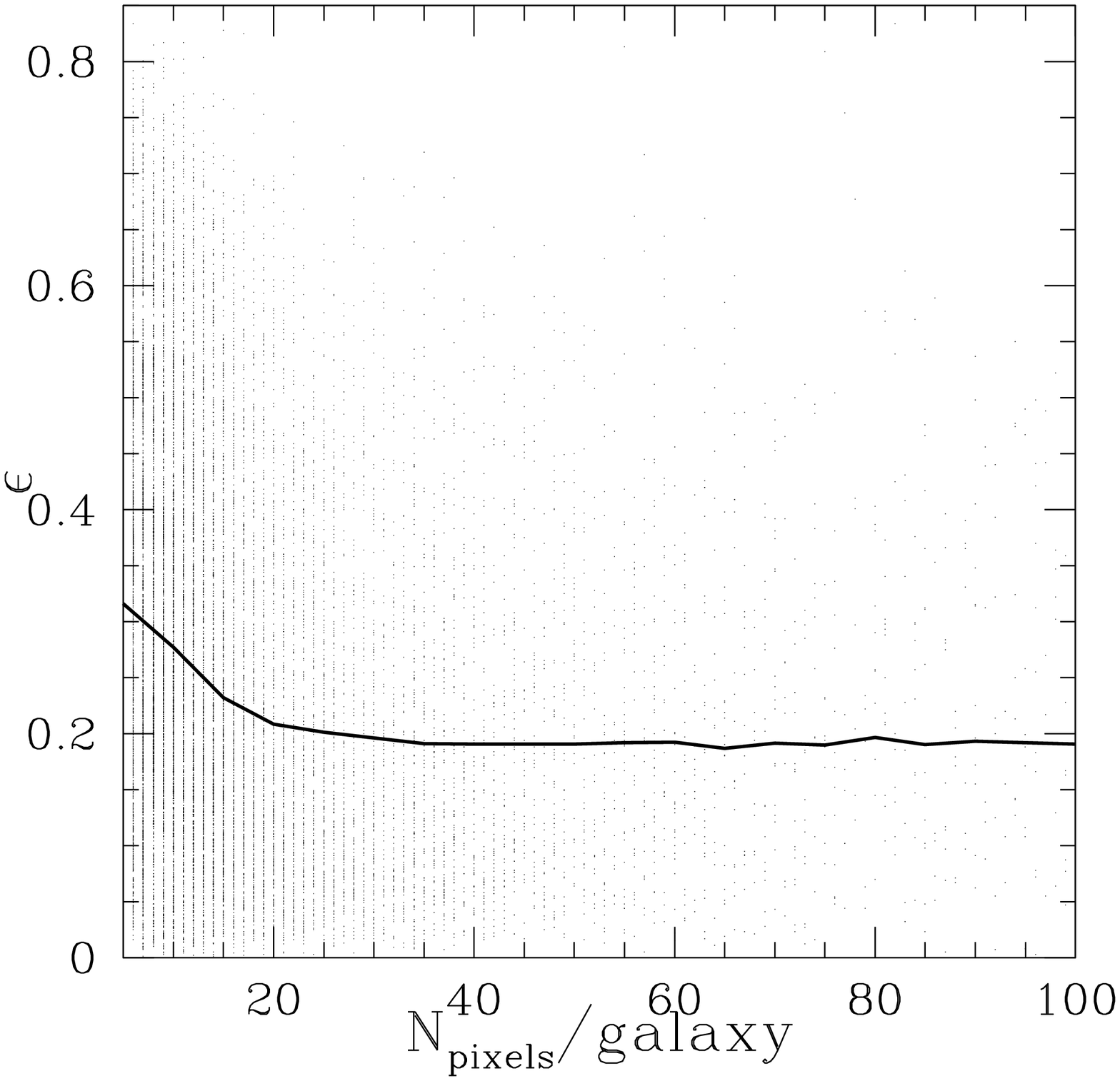}
\caption{Galaxy ellipticities versus galaxy area ($N_{\rm pix}$), 
used to determine shape parameters. The continuous line represents the mean 
ellipticity. In order to visually enhance the problematic 
region of small galaxy areas, we limit the x-axis to 100 pixels/galaxy.}
\end{figure}

\begin{figure}
\plotone{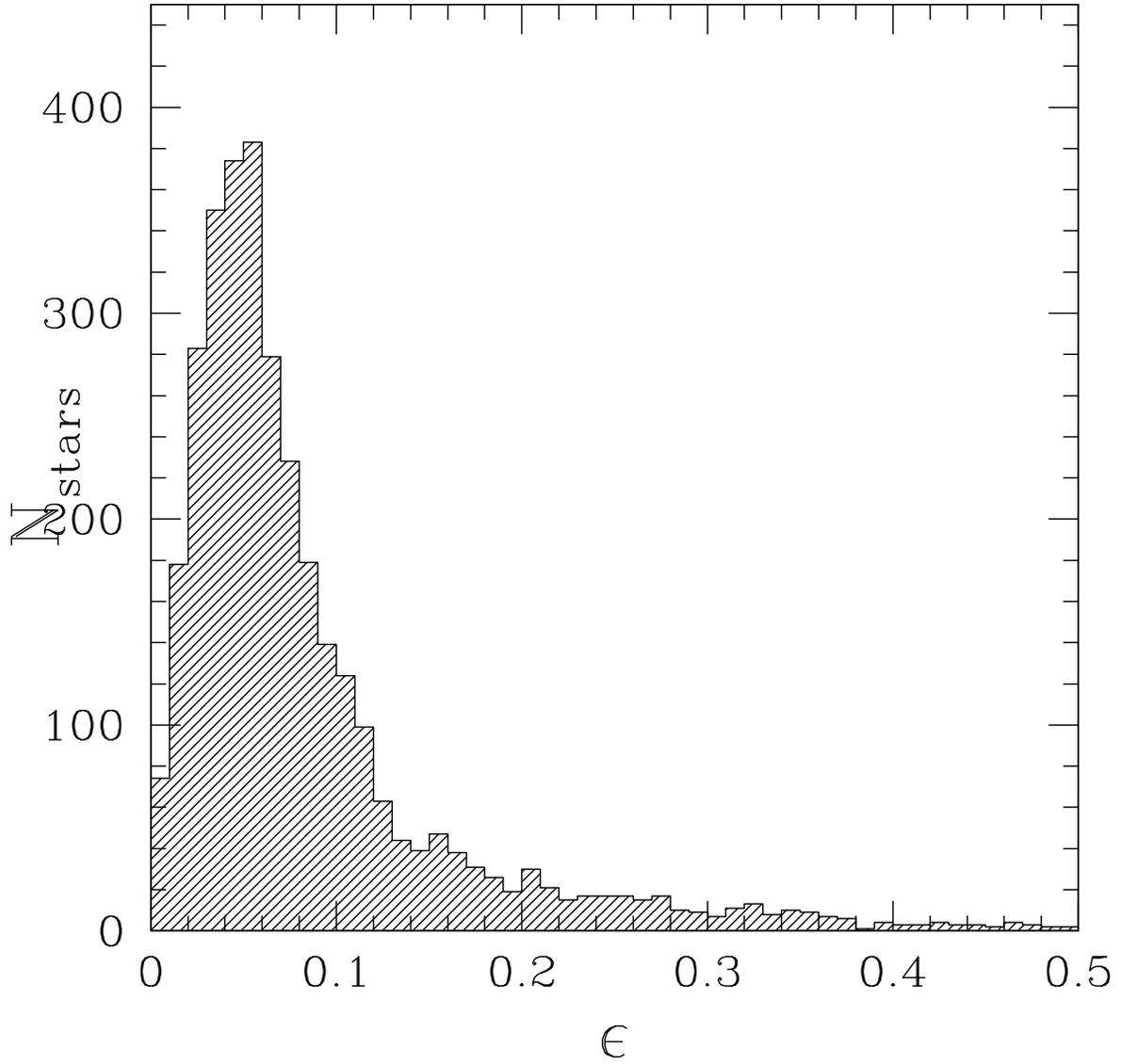}
\caption{The frequency distribution of the non saturated
stellar ellipticities in our
Abell cluster fields. 
%In the insert we show the corresponding
%position angle distribution for all the stars (continuous line) and
%for those with $\epsilon >0.1$ (broken line).
}
\end{figure}

%\begin{figure}
%\plotone{Nfields.ps}
%\caption{The distribution of the number of galaxies per cluster field
%which were used in the analysis.}
%\end{figure}

\begin{figure}
\plotone{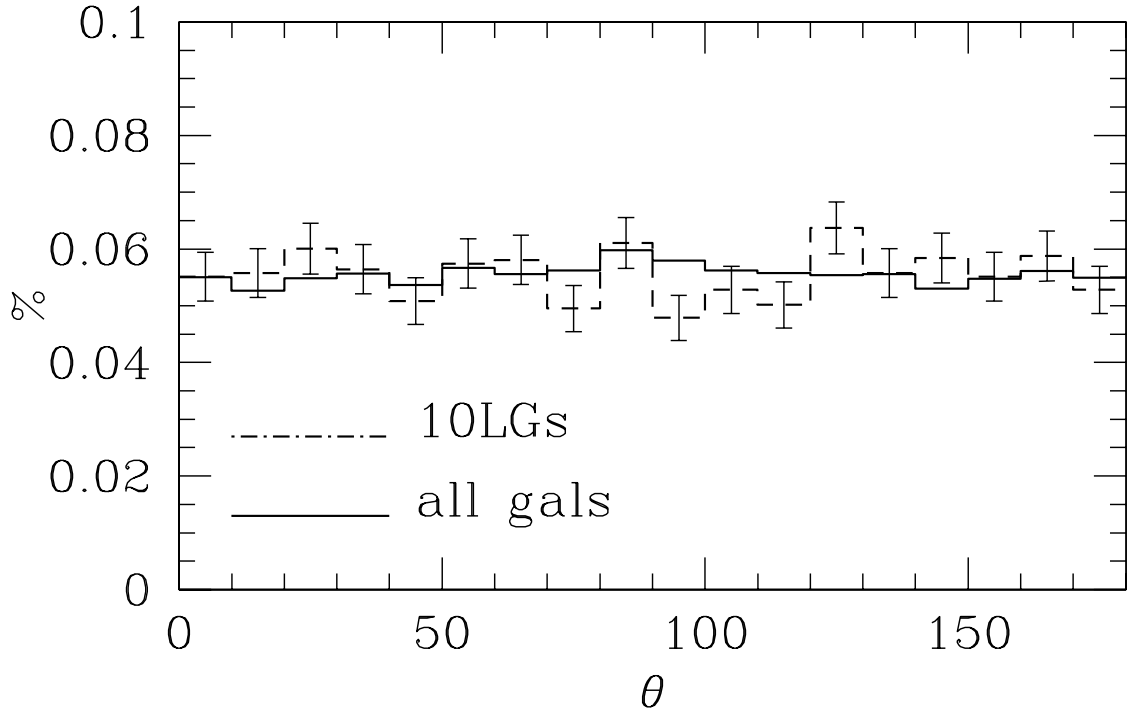}
\caption{The distribution of galaxy position angles for the whole
sample of 15560 galaxies (continuous line) and for the 10 largest
galaxies in each cluster field}
\end{figure}

\begin{figure}
\plotone{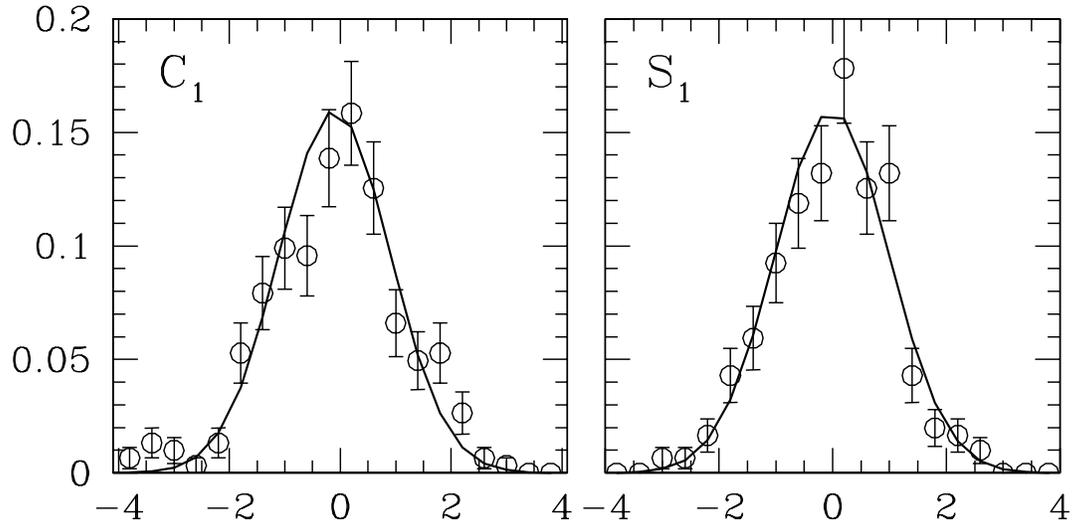}
\caption{The distribution of the first Fourier moments of the galaxy
position angles (points) and the Gaussian expectation in the case of
no systematic directional bias.}
\end{figure}

\begin{figure}
\plotone{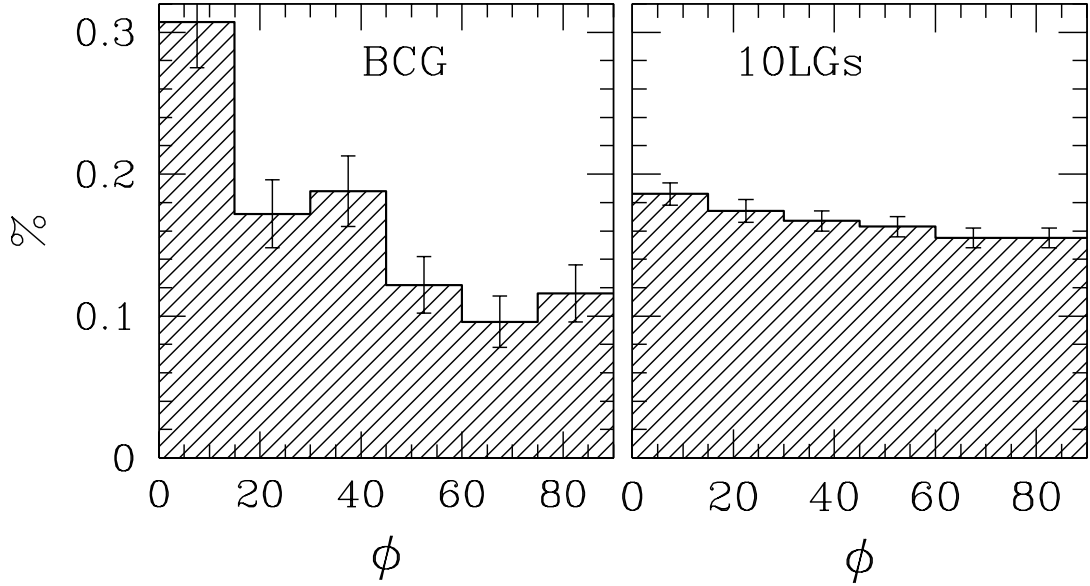}
\caption{The distribution of misalignment angles $\phi$
between the galaxy and cluster orientations. Left panel: BCG
sample. Right Panel: 10LG sample}
\end{figure}

\begin{figure}
\plotone{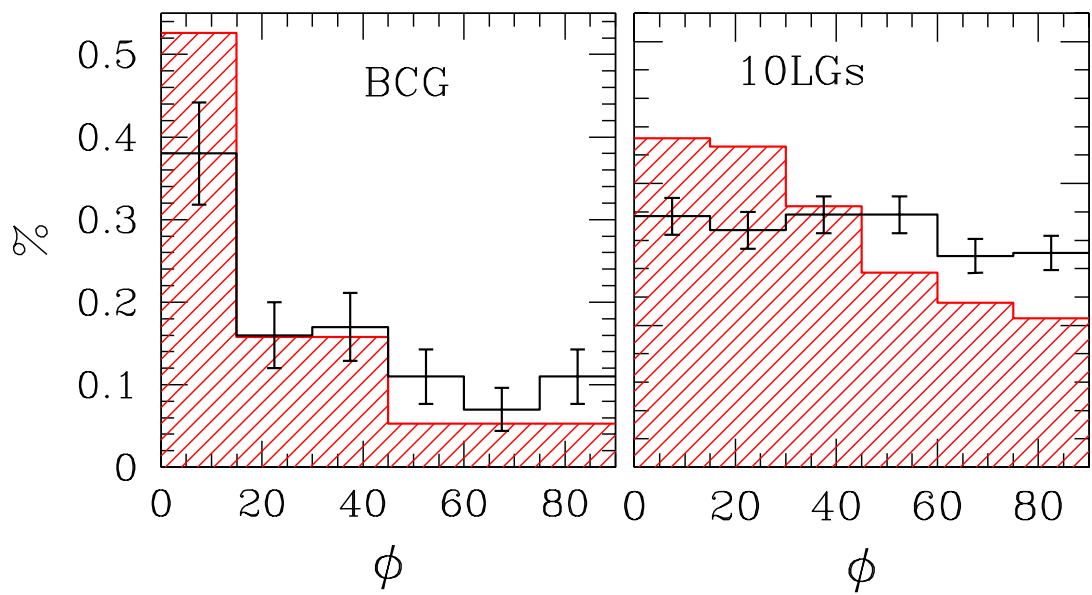}
\caption{The distribution of misalignment angles $\phi$
between the galaxy and cluster orientations for the $\sigma<900$ km/sec 
clusters (black lines) and for those with $\sigma \ge 900$ km/sec
 and found in superclusters 
with $r_{\rm perc}=15$ $h^{-1}$ Mpc (red hatched histogram).}
\end{figure}

\begin{figure}
\plotone{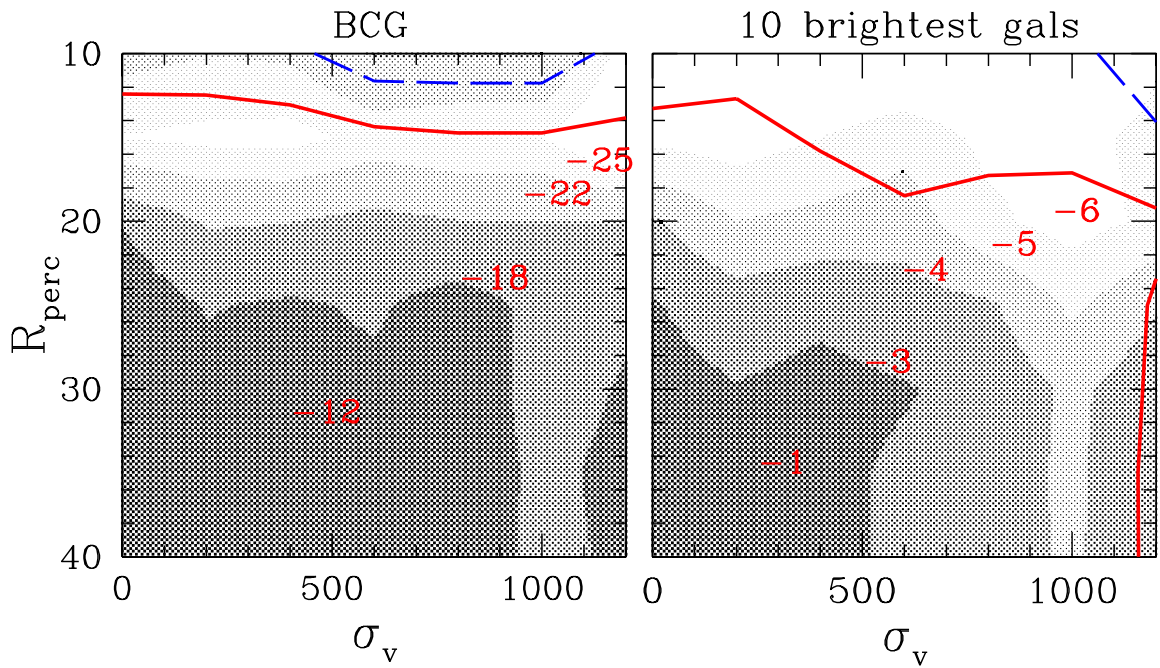}
\caption{Grey-scale plot of the alignment signal, $\langle \delta
\rangle$, indicated by its value, as a function of
membership in superclusters, of the indicated percolation radius,
(y-axis) and of the cluster velocity dispersion (x-axis). 
The contours indicate the significance level at the 3 (continuous
line) and 2.5 $\sigma$ (broken line) levels respectively.}
\end{figure}

\begin{figure}
\plotone{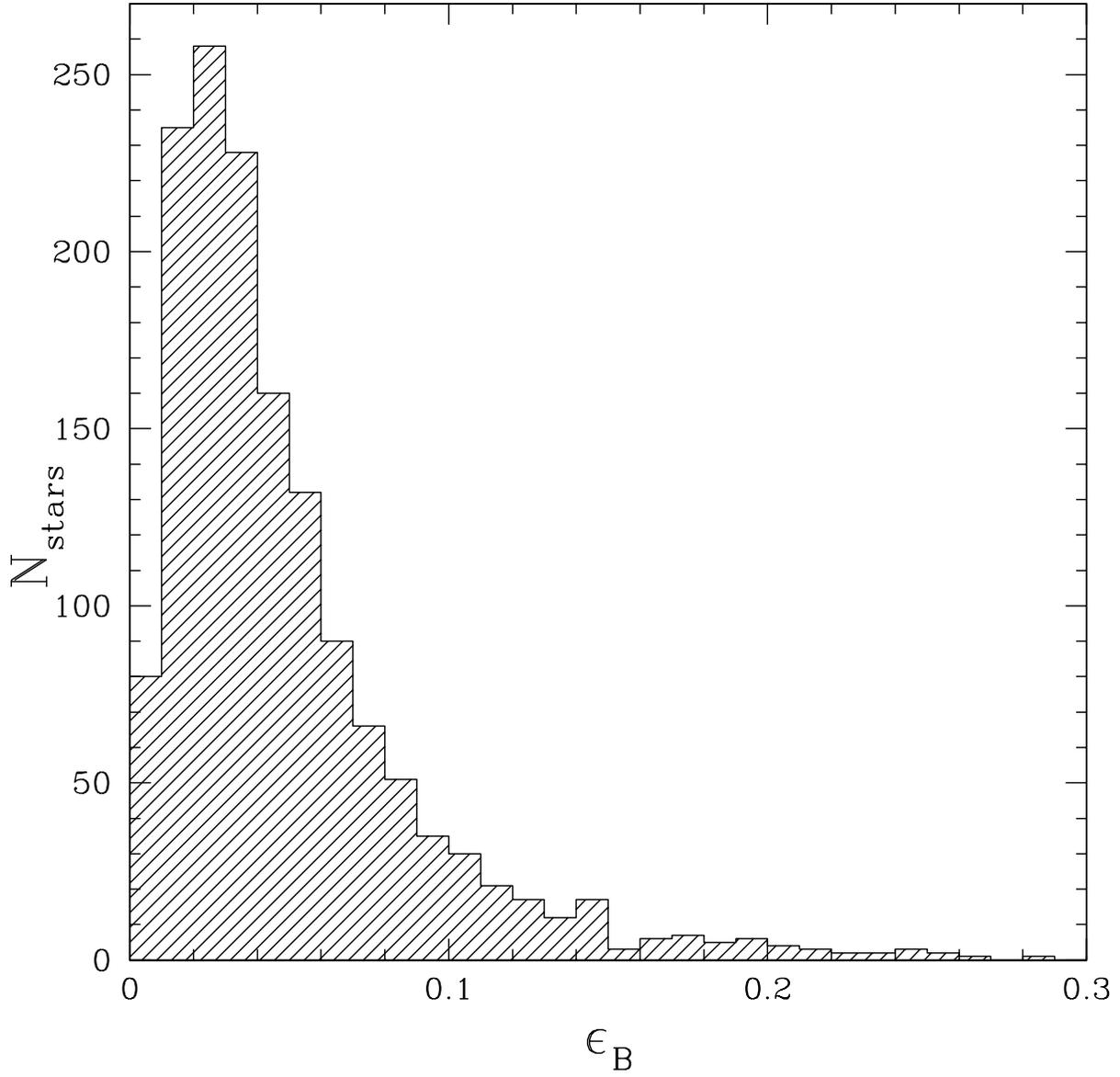}
\caption{The ellipticity frequency distribution in the A521 cluster field.}
\end{figure}

\begin{figure}
\plotone{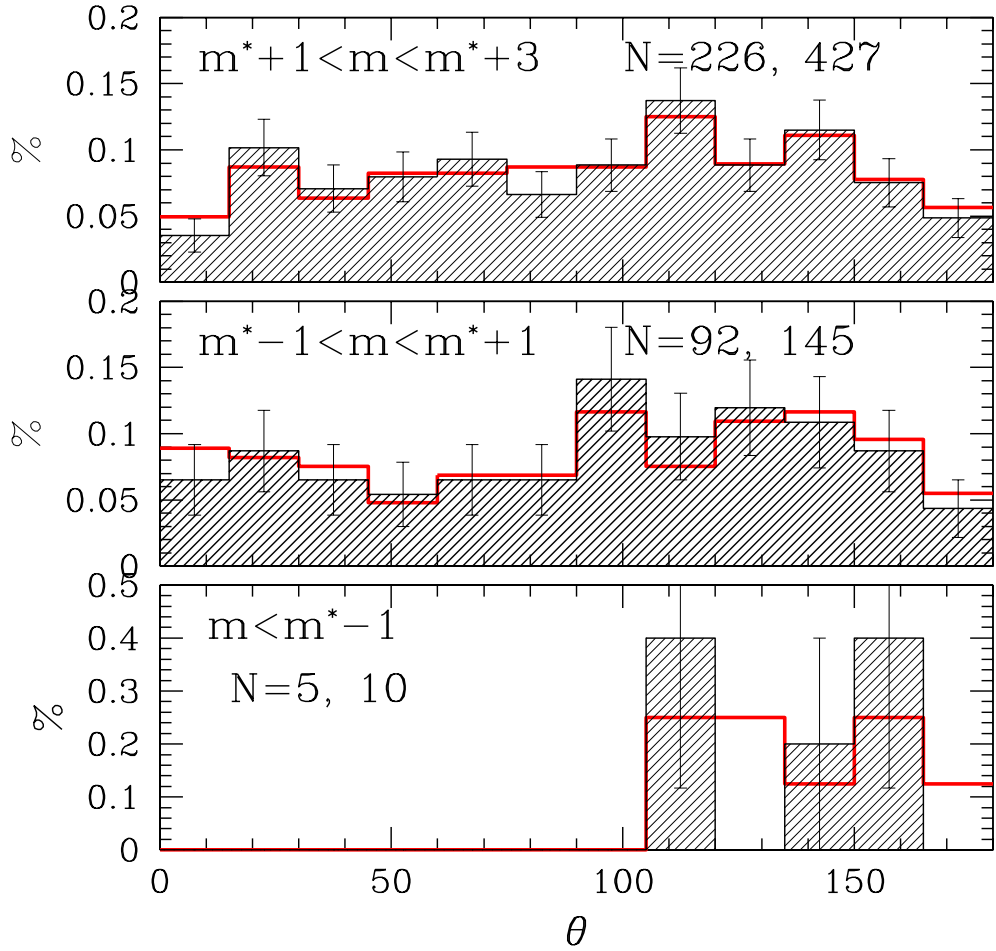}
\caption{Abell 521 galaxy position angle distribution ($\epsilon>0.1$)
for different magnitude limits within 1.5 and 2.5 $h^{-1}$ Mpc
(shaded histogram and thick red line, respectively).}
\end{figure}

\begin{figure}
\plotone{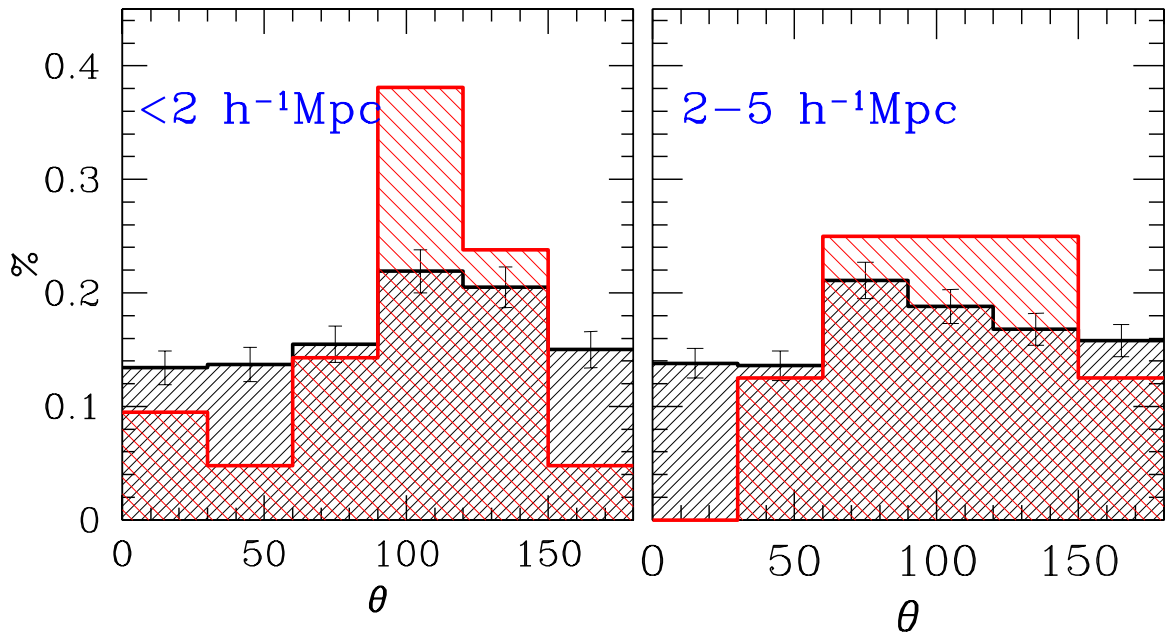}
\caption{Histogram of galaxy (black lines) and groups
(red lines) position angle distribution in Abell 521. 
Left panel: The $ 0 - 2 \;
h^{-1}$ Mpc shell ($m<24.8$), Right panel: The $ 2 - 5 \;
h^{-1}$ Mpc shell ($m<23.8$).}
\end{figure}

\newpage

\begin{table}
\caption[]{Statistical tests of isotropy of the
galaxy position angle distribution for the 303 Abell cluster fields analysed.}
\tabcolsep 14pt
\begin{tabular}{ccccccccc} \\ \hline \hline
sample &  $N_{gals}$ & $C_1$ & $S_1$ & $C_2$ & $S_2$ & ${\cal P}_{R^{2}}$ & 
${\cal P}_{>\chi^2}$ & ${\cal P}_{KS}$ \\ \hline
all gal's & 15560 &-2.3 & 0.1 & 1.6 & -0.8 & 0.08 & 0.31 & 0.78\\
10LG's & 3030 &0.9 & -0.3 & -1.1 & -0.2 & 0.72 & 0.83 & 0.46 
\\ \hline \hline

\end{tabular}
\end{table}

\begin{table}
\caption[]{Galaxy-Cluster Alignment signal and its significance 
for the different samples.}
\tabcolsep 12pt
\begin{tabular}{ccccccc} \\ \hline \hline
sample &  $N_{gals}$ &$\delta$ & $\sigma$ & $S/N$ & ${\cal P}_{>\chi^2}$ & 
${\cal P}_{KS}$ \\ \hline
all gal's & 15560 &-0.9 & 0.2 & 4.5 & 0.035 & 0.005 \\
10LGs & 3030 & -1.6 & 0.5 & 3.3 & 0.028 & 0.003 \\
BCG & 303 & -9.6 & 1.5 & 6.4 & $<10^{-8}$ & $<10^{-8}$ \\
all but BCG \& 10LG's & 12227 &-0.8 & 0.2 & 3.4 & 0.86 & 0.59 
\\ \hline \hline

\end{tabular}
\end{table}

\begin{table}
\caption[]{Abell 521 galaxy major-axes preferred orientations 
and alignment signal 
(within 1.5 or 2.5 $h^{-1}$ Mpc from cluster centre) 
in different magnitude bins. Note that the $B$ magnitude completeness
limit is 24 at 5$\sigma$.}
\tabcolsep 12pt
\begin{tabular}{llcccccc} \\ \hline \hline
mag. bin &  $N_{gals}$ & $\theta_{\rm pr}$ & $P_{>\chi^2}$ & $P_{KS}$ &
$P_{R_2}$ & $\delta$ & $\sigma$ \\ \hline 
\hline \\

  & \multicolumn{7}{c}{$\epsilon>0.1$ and $r<1.5 \; h^{-1}$ Mpc} \\ \hline \\
$<$ 19.8 & 5 & 134$^{\circ}$ & $<10^{-4}$ & $0.02$ & $0.02$ & -29.2 & 11.62\\ 
19.8 - 21.8 & 92 & 118$^{\circ}$ & $0.50$ & $0.16$ & $0.05$ & -4.33 & 2.71\\ 
21.8 - 23.8 & 226 & 109$^{\circ}$ & $0.03$ & $0.08$ &  $10^{-3}$
&-2.26 & 1.73\\
$<$ 24.8 & 435 & 114$^{\circ}$ & $0.002$ & $8\times 10^{-4}$ &  $<10^{-4}$ 
       & -3.5 & 1.24 \\ \hline \\ 

  & \multicolumn{7}{c}{$\epsilon>0.25$ and $r<1.5 \; h^{-1}$ Mpc} \\ \hline \\
$<$ 19.8 & 3 & 148$^{\circ}$ & $<10^{-4}$ & $0.03$ & $0.05$ & -37.0& 15 \\ 
19.8 - 21.8 & 43 & 124$^{\circ}$ & $0.089$ & $0.031$ & $0.02$ & -8.2 & 3.96 \\ 
21.8 - 23.8 & 79 & 104$^{\circ}$ & $0.018$ & $0.45$ &  $0.05$ & -1.83 & 2.92\\ 
$<$ 24.8 & 175 & 115$^{\circ}$ & $0.005$ & $0.002$ &  $3\times 10^{-4}$ 
       & -4.8 & 1.96 \\ \hline \\

%  & \multicolumn{7}{c}{$\epsilon>0.1$, only measured $z$} \\ \hline \\ 
%0 - 20.3 & 46 & 105$^{\circ}$ & $<10^{-4}$ & $0.04$ & $0.001$ & -4.2 & 3.8 \\ 
%\hline \\
  & \multicolumn{7}{c}{$\epsilon>0.1$ and $r<2.5 \; h^{-1}$ Mpc} \\ 
\hline 
$<$ 19.8    & 10  & 136$^{\circ}$ & $<10^{-4}$ & $0.006$   & $0.01$ & -23.0 & 8.2 \\ 
19.8 - 21.8 & 145 & 133$^{\circ}$ & $0.57$     & $0.43$    & $0.10$ & -4.3 & 2.1 \\ 
21.8 - 23.8 & 427 & 109$^{\circ}$ & $0.033$    & $0.062$   & $<10^{-4}$ & -2.2 & 1.2 \\ 
$<$ 24.8    & 804 & 114$^{\circ}$ & $<10^{-4}$ & $<10^{-4}$& $<10^{-4}$ & -3 & 0.9 \\ 
\hline \hline
\end{tabular}
\end{table}

\begin{table}
\caption[]{Preferred Orientation of the Abell 521 galaxy and groups major axis 
distribution for $m_b<24.8$. $P(<\delta\theta)$ is the probability
that their misalignment angle is due to chance.} 
\tabcolsep 15pt
\begin{tabular}{ccccccc} \\ \hline \hline
shell ($h^{-1}$ Mpc) & $N_{\rm gal}$ & $\theta_{\rm pr,gal}$ &
$N_{\rm group}$ & $\theta_{\rm pr,group}$ & $\delta\theta$ &
$P(<\delta\theta)$ \\ \hline
0 - 1 & 237 & 117$^{\circ}$ & 8 & 115$^{\circ}$ & 2$^{\circ}$ &0.0003 \\
1 - 2 & 369 & 113$^{\circ}$ & 7 & 118$^{\circ}$ & 5$^{\circ}$ &0.0019 \\
2 - 3 & 361 & 112$^{\circ}$ & 2 & 144$^{\circ}$ & 32$^{\circ}$ &0.0760 \\
3 - 4 & 442 & 81$^{\circ}$  & 7 & 130$^{\circ}$ & 49$^{\circ}$ &0.1719 \\
4 - 5 & 380 & 113$^{\circ}$ & 4 & 111$^{\circ}$ & 2$^{\circ}$ &0.0003
\\ \hline \hline

\end{tabular}
\end{table}


\begin{thebibliography}{}
\bibitem{} Abell, G.O., 1958, ApJS, 3, 211
\bibitem{}Abell, G.O., Corwin, H.G., Jr., Olowin, R.P., 1989, ApJS,
70, 1
\bibitem{} Adams, M.T., Strom, K.M., Strom, S.E., 1980, ApJ, 238, 445
\bibitem{}Arnaud, M., Maurogordato,S., Slezak, E., Rho,J., 2000, A\&A, 355, 461
\bibitem{}Barnes, J. \& Efstathiou, G., 1987, ApJ, 319, 575
\bibitem{} Bartelmann, M., Schneider, P., 2001, Phys.Rep., 340, 291
\bibitem{} Basilakos, S., Plionis, M., Maddox, S., 2000, \mnras, 316, 779
\bibitem{} Bertin, E.; Arnouts, A\&AS, 1996, 117, 393
\bibitem{}Bingelli B., 1982, AA, 250, 432
\bibitem{} Brown, M.N., Taylor, A.N., Hambly, N.C., Dye, S., 2002,
MNRAS, 333, 501
\bibitem{} Carter, D. \& Metcalfe, N., 1980, \mnras, 191, 325
\bibitem{} Catelan, P., Kamionkowski, M., Blandford, R.D., 2001,
  MNRAS, 320, L7
\bibitem{}Chambers, S.W., Melott, A.L., Miller, C.J., 2002, \apj, 565, 849
\bibitem{} Bond, J.R., 1987, in {\em Nearly Normal Galaxies}, 
ed. Faber, S., (New York: Springer-Verlag), p.388 (Dordrecht: Reidel), p.255
\bibitem{} Cabanela, J. E.; Aldering, G., 1998, AJ, 116, 1094
\bibitem{}Coutts, A., 1996, MNRAS, 278, 87
\bibitem{}Djorgovski, S., 1983, ApJ, 274, L7
\bibitem{}Djorgovski, S., 1987, in {\em Nearly Normal Galaxies}, p.227, 
Springer-Verlag
\bibitem{}Durret, F., Forman, W., Gerbal, D., Jones, C., Vikhlinin,
A., 1998, A\&A, 335, 41
\bibitem{} Faltenbacher, A., Kerscher, M., Gottloeber,S., Mueller,
M., 2002, A\&A, 395, 1
\bibitem{}Feretti, L., 2001, in IAU Symp. 199, The Universe at Low Radio 
Frequencies, ed. A. Pramesh Rao (San Francisco: ASP)
\bibitem{}Feretti, L., \& Giovannini, G., 1996, in IAU Symp. 175, 
Extragalactic Radio Sources, ed. R. Ekers, C. Fanti, \& L. Padrielli 
(Dordrecht: Kluwer), 333
\bibitem{} Ferrari, C., Maurogordato, S., Cappi, A., \& Benoist, C., 2003, A\&A, 399, 813
\bibitem{}Fuller, T.M., West, M.J. \& Bridges, T.J., 1999, ApJ, 519, 22 
\bibitem{} van Kampen, E., Rhee, G.F.R.N., 1990, A\&A, 237, 283
\bibitem{} Heymans, C., Heavens, A., 2003, MNRAS, 339, 711
\bibitem{} Kim, R.S.J., et al. (SDSS collaboration), 2001, 
{\em astro-ph/0110383}
\bibitem{} King, L.J., Schneider, P., 2003, A\&A, 398, 23
\bibitem{}Lee, J. \& Pen, U., 2002, ApJ, 567, L111 
\bibitem{} Mardia, K.V., 1972, {\em Statistics of Directional Data},
(Academic, London) 
\bibitem{}Maurogordato, S., Proust, D., Beers, T., Arnaud, M.,Pello, R., 
Capp, A., Slezak, E., 2000, A\&A, 355, 848
\bibitem{} Onuora, L.I., Thomas, P.A, 2000, MNRAS, 319, 614
\bibitem{}Plionis M., 1994, ApJS., 95, 401
\bibitem{}Plionis M., 2001, in the proceedings of the {\em Clusters and 
the High-Redshift Universe observed in X-rays}, XXI$^{\rm th}$ Moriond 
Astrophysics Meeting, eds. Neumann et al., {\em in press}
\bibitem{} Plionis, M. \& Basilakos, S., 2002, \mnras, 329, L47
\bibitem{} de Propris, R. et al.,
2002, \mnras, 329, 227
%\bibitem[e.g., Ritchie \& Thomas(2002)]{Ritchi}Ritchie, B.W., 
%Thomas, P.A., 2002, \mnras, 329, 675
\bibitem{}Quinn, T. \& Binney, J., 1992, MNRAS, 255, 729
\bibitem{}Salvador-Sole, E. \& Solanes, J.M., 1993, ApJ, 417, 427
\bibitem{} Sastry, G.N., 1968, \pasp, 80, 252
\bibitem{}Sch\"{u}ecker, P., B\"{o}hringer, H., Reiprich, T. H., 
Feretti, L., AA, 2001, 378, 408
\bibitem{}Splinter, R.J., Melott, A.L., Linn, A.M., Buck, C.,
 Tinker, J., 1997, ApJ, 479, 632
\bibitem{}Struble, M.F., 1990, AJ, 99, 743
\bibitem{}Struble, M.F., Peebles, P.J.E., 1985, AJ, 90, 582
\bibitem{}Struble, M.F., Rood, H.J., 1999, ApJS, 125, 35
\bibitem{}Usami, M. \& Fujimoto, M., 1997, ApJ, 487, 489
\bibitem{}Tormen, G., 1997, MNRAS, 290, 411
\bibitem{}Trevese, D., Cirimele, G., Flin, P., 1992, AJ, 104, 935
\bibitem{}van Haarlem, M., van de Weygaert, R., 1993, ApJ, 418, 544
\bibitem{}West, M. J., 1989, ApJ, 347, 610
\bibitem{}West, M. J., Villumsen, J.V., Dekel, A., 1991, ApJ, 369, 287
\bibitem{}West, M. J., 1994, MNRAS, 268, 79
\bibitem[e.g. West, Jones \& Forman(1995)]{West} West, M.J., Jones, C.,
\& Forman, W., 1995, \apj, 451, L5
\end{thebibliography}
\end{document}